\documentclass[12pt]{article}
\usepackage{epsfig}
\usepackage{ifthen} 

\newboolean{uprightparticles}
\setboolean{uprightparticles}{false} 

\usepackage{microtype}
\usepackage{lineno}  
\usepackage{xspace} 

\usepackage{graphicx}  
\usepackage{color}
\usepackage{colortbl}
\graphicspath{{./figs/}} 

\usepackage{amsmath} 
\usepackage{amssymb}
\usepackage{amsfonts}
\usepackage{upgreek} 

\newcommand*\patchAmsMathEnvironmentForLineno[1]{%
\expandafter\let\csname old#1\expandafter\endcsname\csname #1\endcsname
\expandafter\let\csname oldend#1\expandafter\endcsname\csname
end#1\endcsname
 \renewenvironment{#1}%
   {\linenomath\csname old#1\endcsname}%
   {\csname oldend#1\endcsname\endlinenomath}%
}
\newcommand*\patchBothAmsMathEnvironmentsForLineno[1]{%
  \patchAmsMathEnvironmentForLineno{#1}%
  \patchAmsMathEnvironmentForLineno{#1*}%
}
\AtBeginDocument{%
\patchBothAmsMathEnvironmentsForLineno{equation}%
\patchBothAmsMathEnvironmentsForLineno{align}%
\patchBothAmsMathEnvironmentsForLineno{flalign}%
\patchBothAmsMathEnvironmentsForLineno{alignat}%
\patchBothAmsMathEnvironmentsForLineno{gather}%
\patchBothAmsMathEnvironmentsForLineno{multline}%
}

\def\ux85 {\mbox{UX85}\xspace}

\ifthenelse{\boolean{uprightparticles}}%
{

 \def\Ppi         {\ensuremath{\uppi}\xspace}

 \def\Ppsi        {\ensuremath{\uppsi}\xspace}

 \def\PDelta      {\ensuremath{\Delta}\xspace}                 
 \def\PXi      {\ensuremath{\Xi}\xspace}                 
 \def\PLambda      {\ensuremath{\Lambda}\xspace}                 
 \def\PSigma      {\ensuremath{\Sigma}\xspace}                 
 \def\POmega      {\ensuremath{\Omega}\xspace}                 
 \def\PUpsilon      {\ensuremath{\Upsilon}\xspace}                 
 
 \def\PB      {\ensuremath{\mathrm{B}}\xspace}                 
                  
 \def\PD      {\ensuremath{\mathrm{D}}\xspace}

 \def\PJ      {\ensuremath{\mathrm{J}}\xspace}                 
 \def\PK      {\ensuremath{\mathrm{K}}\xspace}

 \def\Pb      {\ensuremath{\mathrm{b}}\xspace}

 \def\Pi      {\ensuremath{\mathrm{i}}\xspace}

 \def\Ps      {\ensuremath{\mathrm{s}}\xspace}

}
{

 \def\Ppi         {\ensuremath{\pi}\xspace}

 \def\Ppsi        {\ensuremath{\psi}\xspace}                 
                  
 \mathchardef\PDelta="7101
 \mathchardef\PXi="7104
 \mathchardef\PLambda="7103
 \mathchardef\PSigma="7106
 \mathchardef\POmega="710A
 \mathchardef\PUpsilon="7107
                  
 \def\PB      {\ensuremath{B}\xspace}                 
                  
 \def\PD      {\ensuremath{D}\xspace}

 \def\PJ      {\ensuremath{J}\xspace}                 
 \def\PK      {\ensuremath{K}\xspace}

 \def\Pb      {\ensuremath{b}\xspace}

 \def\Pi      {\ensuremath{i}\xspace}

 \def\Ps      {\ensuremath{s}\xspace}

}

\def\squark    {\ensuremath{\Ps}\xspace}

\def\bquark    {\ensuremath{\Pb}\xspace}

\def\pion  {\ensuremath{\Ppi}\xspace}

\def\pip   {\ensuremath{\pion^+}\xspace}
\def\pim   {\ensuremath{\pion^-}\xspace}

\def\pipm  {\ensuremath{\pion^\pm}\xspace}
\def\pimp  {\ensuremath{\pion^\mp}\xspace}

\def\kaon  {\ensuremath{\PK}\xspace}
  \def\Kbar  {\kern 0.2em\overline{\kern -0.2em \PK}{}\xspace}

\def\Kz    {\ensuremath{\kaon^0}\xspace}
\def\Kzb   {\ensuremath{\Kbar^0}\xspace}
\def\KzKzb {\ensuremath{\Kz \kern -0.16em \Kzb}\xspace}
\def\Kp    {\ensuremath{\kaon^+}\xspace}
\def\Km    {\ensuremath{\kaon^-}\xspace}
\def\Kpm   {\ensuremath{\kaon^\pm}\xspace}

\def\KpKm  {\ensuremath{\Kp \kern -0.16em \Km}\xspace}

  \def\Dbar    {\kern 0.2em\overline{\kern -0.2em \PD}{}\xspace}
\def\D       {\ensuremath{\PD}\xspace}

\def\Dz      {\ensuremath{\D^0}\xspace}
\def\Dzb     {\ensuremath{\Dbar^0}\xspace}
\def\DzDzb   {\ensuremath{\Dz {\kern -0.16em \Dzb}}\xspace}
\def\Dp      {\ensuremath{\D^+}\xspace}
\def\Dm      {\ensuremath{\D^-}\xspace}

\def\DpDm    {\ensuremath{\Dp {\kern -0.16em \Dm}}\xspace}

\def\B       {\ensuremath{\PB}\xspace}
  \def\Bbar    {\kern 0.18em\overline{\kern -0.18em \PB}{}\xspace}

\def\Bz      {\ensuremath{\B^0}\xspace}

\def\Bu      {\ensuremath{\B^+}\xspace}
\def\Bub     {\ensuremath{\B^-}\xspace}
\def\Bp      {\ensuremath{\Bu}\xspace}
\def\Bm      {\ensuremath{\Bub}\xspace}
\def\Bpm     {\ensuremath{\B^\pm}\xspace}

\def\Bd      {\ensuremath{\B^0}\xspace}
\def\Bs      {\ensuremath{\B^0_\squark}\xspace}

\def\jpsi     {\ensuremath{{\PJ\mskip -3mu/\mskip -2mu\Ppsi\mskip 2mu}}\xspace}

  \def\Y#1S{\ensuremath{\PUpsilon{(#1S)}}\xspace}

\def\Lbar {\ensuremath{\kern 0.1em\overline{\kern -0.1em\PLambda}}\xspace}

\def\to                 {\ensuremath{\rightarrow}\xspace}

\def\CP                {\ensuremath{C\!P}\xspace}

\def\AT#1     {\ensuremath{A_{\mathrm{T}}^{#1}}\xspace}           

\def\C#1      {\ensuremath{\mathcal{C}_{#1}}\xspace}                       
\def\Cp#1     {\ensuremath{\mathcal{C}_{#1}^{'}}\xspace}                    
\def\Ceff#1   {\ensuremath{\mathcal{C}_{#1}^{\mathrm{(eff)}}}\xspace}        
\def\Cpeff#1  {\ensuremath{\mathcal{C}_{#1}^{'\mathrm{(eff)}}}\xspace}       
\def\Ope#1    {\ensuremath{\mathcal{O}_{#1}}\xspace}                       
\def\Opep#1   {\ensuremath{\mathcal{O}_{#1}^{'}}\xspace}                    

\newcommand{\tev}{\ensuremath{\mathrm{\,Te\kern -0.1em V}}\xspace}
\newcommand{\gev}{\ensuremath{\mathrm{\,Ge\kern -0.1em V}}\xspace}
\newcommand{\mev}{\ensuremath{\mathrm{\,Me\kern -0.1em V}}\xspace}
\newcommand{\kev}{\ensuremath{\mathrm{\,ke\kern -0.1em V}}\xspace}
\newcommand{\ev}{\ensuremath{\mathrm{\,e\kern -0.1em V}}\xspace}
\newcommand{\gevc}{\ensuremath{{\mathrm{\,Ge\kern -0.1em V\!/}c}}\xspace}
\newcommand{\mevc}{\ensuremath{{\mathrm{\,Me\kern -0.1em V\!/}c}}\xspace}
\newcommand{\gevcc}{\ensuremath{{\mathrm{\,Ge\kern -0.1em V\!/}c^2}}\xspace}
\newcommand{\gevgevcccc}{\ensuremath{{\mathrm{\,Ge\kern -0.1em V^2\!/}c^4}}\xspace}
\newcommand{\mevcc}{\ensuremath{{\mathrm{\,Me\kern -0.1em V\!/}c^2}}\xspace}

\newcommand{\stat}{\ensuremath{\mathrm{(stat)}}\xspace}
\newcommand{\syst}{\ensuremath{\mathrm{(syst)}}\xspace}

\def\gsim{{~\raise.15em\hbox{$>$}\kern-.85em
          \lower.35em\hbox{$\sim$}~}\xspace}
\def\lsim{{~\raise.15em\hbox{$<$}\kern-.85em
          \lower.35em\hbox{$\sim$}~}\xspace}

\def\sqs   {\ensuremath{\protect\sqrt{s}}\xspace}

\def\tell1  {TELL1\xspace}
\def\ukl1   {UKL1\xspace}

\usepackage{cite} 
\usepackage{mciteplus}

\def\pipipi {\ensuremath{\Bpm \to \pipm \pip \pim}\xspace}
\def\kpipi {\ensuremath{\Bpm \to \Kpm \pip \pim}\xspace}
\def\kkpi {\ensuremath{\Bpm \to \Kp \Km \pipm}\xspace}
\def\kkk {\ensuremath{\Bpm \to \Kpm \Kp \Km}\xspace}

\def\jpsik {\ensuremath{\Bpm \to \jpsi \Kpm}\xspace}

\def\fspipipi {\ensuremath{\pipm \pip \pim}\xspace}
\def\fskpipi {\ensuremath{\Kpm \pip \pim}\xspace}
\def\fskkpi {\ensuremath{\Kp \Km \pipm}\xspace}
\def\fskkk {\ensuremath{\Kpm \Kp \Km}\xspace}
\def\fsjpsik {\ensuremath{\jpsi \Kpm}\xspace}

\def\kpipip {\ensuremath{\Bp \to \Kp \pip \pim}\xspace}

\def\kkkp {\ensuremath{\Bp \to \Kp \Kp \Km}\xspace}

\def\kkpip {\ensuremath{\Bp \to \Kp \Km \pip}\xspace}

\def\pipipip {\ensuremath{\Bp \to \pip \pip \pim}\xspace}

\def\Bsz {\ensuremath{\Bs  \to K \pi}\xspace}
\def\Bz {\ensuremath{\Bd  \to K \pi}\xspace}
\def\BszC {\ensuremath{\Bs  \to \Km \pip}\xspace}
\def\BzC {\ensuremath{\Bd  \to \Kp \pim}\xspace}

\def\mmkpi {\ensuremath{m^2_{\Kpm \pimp}}\xspace}                    
\def\mmpipi {\ensuremath{m^2_{\pip \pim}}\xspace}                        
\def\mmkk {\ensuremath{m^2_{\Kp \Km}}\xspace}                            
\def\mmkklow {\ensuremath{m^2_{\Kp \Km\,{\rm low}}}\xspace}     
\def\mmpipilow {\ensuremath{m^2_{\pip \pim\,{\rm low}}}\xspace} 
\def\mmpipihi {\ensuremath{m^2_{\pip \pim\,{\rm high}}}\xspace}   

\def\acp {\ensuremath{A_{\CP}}\xspace}
\def\acpraw {\ensuremath{A_{\CP}^{\rm RAW}}\xspace}
\def\acprawacc {\ensuremath{{^{\rm{ACC}}}\!\!A_{\CP}^{\rm RAW}}\xspace}
\def\acpn {\ensuremath{A_{\CP}^{N}}\xspace}

\def\adetk {\ensuremath{A_{\rm D}^{K}}\xspace}
\def\adetpi {\ensuremath{A_{\rm D}^{\pi}}\xspace}

\def\valbmkkpi {\ensuremath{619}\xspace}
\def\errbmkkpi {\ensuremath{47}\xspace}
\def\valbpkkpi {\ensuremath{875}\xspace}
\def\errbpkkpi {\ensuremath{50}\xspace}

\def\valbmpipipi {\ensuremath{2718}\xspace}
\def\errbmpipipi {\ensuremath{71}\xspace}
\def\valbppipipi {\ensuremath{2111}\xspace}
\def\errbppipipi {\ensuremath{66}\xspace}

\def\erracpjpsikpdg {\ensuremath{0.007}\xspace}

\def\valacpkkpiregAcc {\ensuremath{-0.671}\xspace}
\def\erracpkkpiregAcc {\ensuremath{0.067}\xspace}

\def\errsystacpkkpiregAcc {\ensuremath{0.028}\xspace} 

\def\valacppipipiregAcc {\ensuremath{0.622}\xspace}
\def\erracppipipiregAcc {\ensuremath{0.075}\xspace}

\def\errsystacppipipiregAcc {\ensuremath{0.032}\xspace} 

\def\valacpkkpiAcc {\ensuremath{-0.153}\xspace}
\def\errstatacpkkpiAcc {\ensuremath{0.046}\xspace}
\def\errsystacpkkpiAcc {\ensuremath{0.019}\xspace} 

\def\valacppipipiAcc {\ensuremath{0.120}\xspace}
\def\errstatacppipipiAcc {\ensuremath{0.020}\xspace}
\def\errsystacppipipiAcc {\ensuremath{0.019}\xspace}

\def\beq{\begin{equation}}
\def\eeq#1{\label{#1}\end{equation}}
\def\eeqn{\end{equation}}

\def\beqa{\begin{eqnarray}}
\def\eeqa#1{\label{#1}\end{eqnarray}}
\def\eeqan{\end{eqnarray}}

\let\bar=\overbar

\def\D{{\cal D}}

\def\Dslash{\not{\hbox{\kern-4pt $D$}}}
\def\dslash{\not{\hbox{\kern-2pt $\del$}}}

\def\msb{{\bar{\ssstyle M \kern -1pt S}}}

\def\Title#1{\begin{center} {\Large {\bf #1} } \end{center}}

\begin{document}
\begin{flushright}
{\small 
Proceedings of CKM 2012, the 7th International Workshop on the CKM Unitarity Triangle, University of Cincinnati, USA, 28 September - 2 October 2012 
}
\end{flushright}

\Title{Direct \CP Violation in Charmless \\B Decays at LHCb}

\bigskip\bigskip


\begin{raggedright}  

{\it Jussara M. de Miranda\index{Miranda , J. M. de}, on behalf of the LHCb collaboration\\
Centro Brasileiro de Pesquisas Fisicas\\
Rio de Janeiro, BRAZIL}
\bigskip\bigskip
\end{raggedright}

\begin{abstract}
  \noindent
Using data collected by LHCb experiment in 2011 we report the measurements of charge asymmetries 
in charmless decays of $B$  mesons in two or three charged kaons or pions. We find positive charge asymmetries in the 
channels \BszC ($3.3\sigma$),  \kpipip ($2.8\sigma$), and \pipipip ($4.2\sigma$)  
and negative in  \BzC ($6\sigma$),\kkkp ($3.7\sigma$),\kkpip ($3.0\sigma$).

\end{abstract}

\section{Introduction}

We present recent studies from the LHCb experiment on direct \CP violation in  $B $ meson
charmless decays. Using data collected in 2011 we have studied the decays:
\BzC and \BszC ~\cite{Aaij:2012qe}; \kpipip and  \kkkp ~\cite{khh} and   \kkpip and \pipipip ~\cite{pihh} 
(complex-conjugate modes are implied except in asymmetry definitions).
The main observable in all the analyses is the charge asymmetry in decay rates into a final state $f$, defined as $\acp={\Gamma(\bar{B}\to
\bar{f})-\Gamma(B\to f)\over \Gamma(\bar{B}\to \bar{f})+\Gamma(B\to f)}$.
These channels have been extensively studied by the \B factories and the Tevatron \cite{Nakamura:2010zzi}
but only now, with the large samples collected by LHCb, it was possible to establish first evidence (significance $>3\sigma$) 
of \CP asymmetry in four of the channels. The exceptions are
the well established \BzC \cite{Nakamura:2010zzi} for which we report on the most precise measurement and the \kpipip
for which our measurement for the inclusive \acp has a significance of 2.8$\sigma$.
The main features of the  LHCb experiment that made this possible are: the very efficient particle identification and vertex location system,
dedicated hadronic trigger and the ability to reverse the magnet polarity on the data taken~\cite{Alves:2008zz}.

\CP violating transitions and possible new physics effects originating in loop diagrams are accessible in all channels.
The decays \kpipip , \kkkp and \BzC proceed through similar \CP conserving penguin diagrams and have access to CKM \CP
 violating transitions $ b \to u$ at tree level. In these cases the smallness of  penguin amplitudes is
 compensated by $\sim\lambda$$^2$ factor as compared to the factor  $\sim\lambda$$^4$ for the
 tree contribution,  where $\lambda\sim0.22$ is the
 expansion factor of the Wolfenstein parametrization of the CKM matrix.  \CP violation effects can potentially be large.
 The decays \kkpip, \pipipip and \BszC access  \CP violating transitions $ b\to u$  and $b \to t$ in
 tree and penguin diagrams respectively, both at the same order  $\sim\lambda$$^3$ leading to
 the overall penguin contribution much smaller than the tree and observation of  \CP violation is disfavored.

Using 0.35~fb$^{-1}$  of $pp$ collisions at  \sqs = 7 TeV, LHCb established the first evidence of \CP violation in the \Bs system in its decays to
$ K\pi$ pairs, 
and also reported  the most precise measurement of the \CP asymmetry of \Bz. 
These results are presented in Section 2.

The three-body decays offer a good environment to study \CP violation enriched by the
 interference pattern of the two-body resonances in the Dalitz plot. The charge asymmetry in regions of the
phase space is likely to be greater than the integrated ones.
 With the complete data set of 1~fb$^{-1}$ collected in 2011, we measure inclusive \CP asymmetries
 for the three-body channels as well as  their distributions  in the phase space.
 It is a consequence of  CPT invariance and unitarity  that the sum of all  partial decay widths of a particle should match that of its antiparticle.
 In fact, CPT is more restrictive. If we divide all possible decays of a particle into classes containing final states that are mutually distinct under
 the strong interactions,  CPT imposes the particle-antiparticle match should be observed 
 in each class\cite{Bigi:2009}. 
Motivated by  the physics, by  CPT invariance that connects the decays \fskpipi to \fskkk and \fskkpi to \fspipipi final states in two distinct classes,
and also due to the experimental resemblance, 
it was natural to group the three-body analysis in two sets: \kpipi and \kkk are described in Section 3 and \kkpi and  \pipipi in Section 4.

\section{Direct \CP violation in \BszC and \BzC decays}

\begin{figure}[htb]
\centering
\includegraphics[width=5.cm]{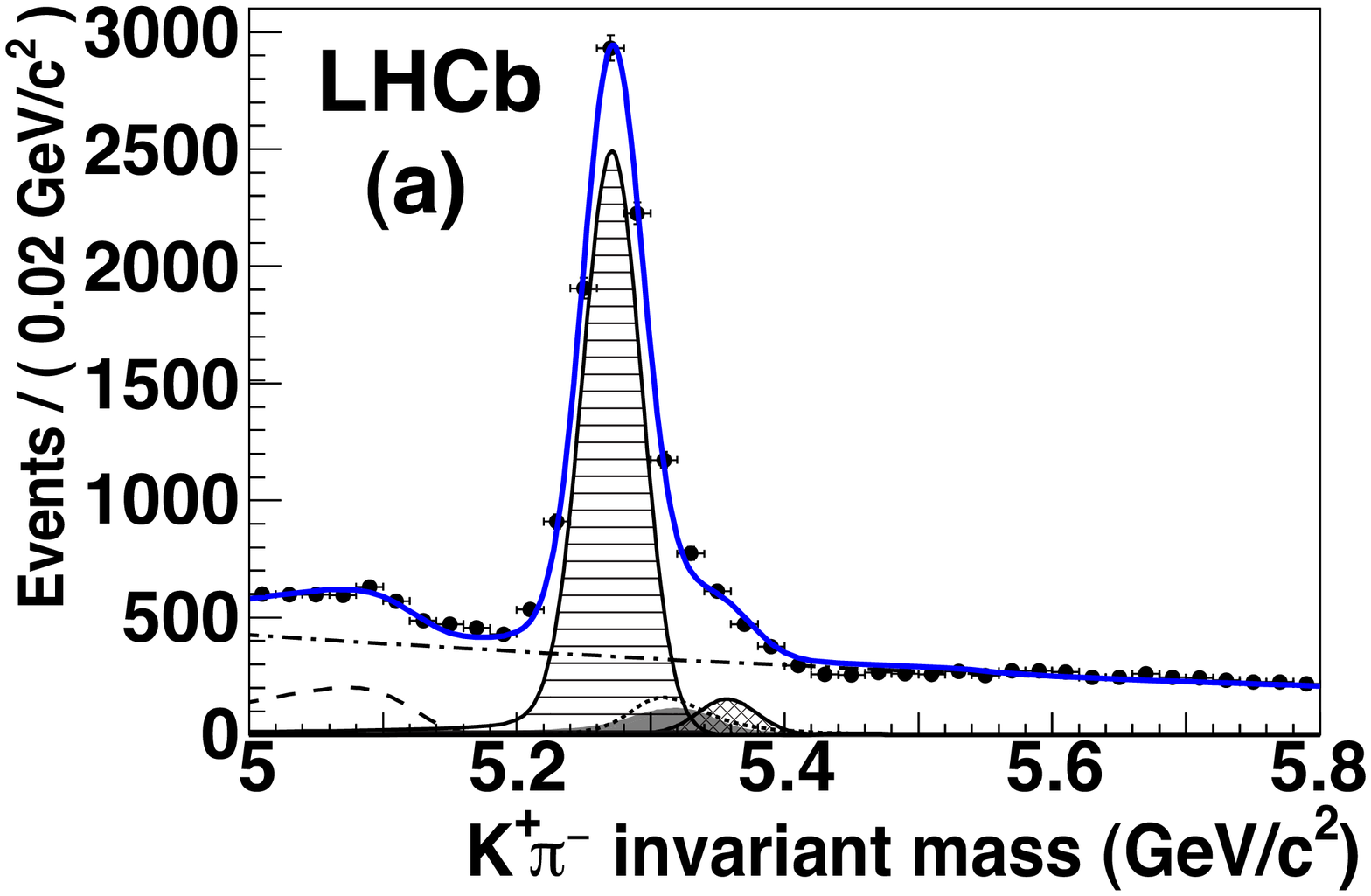}
\includegraphics[width=5.cm]{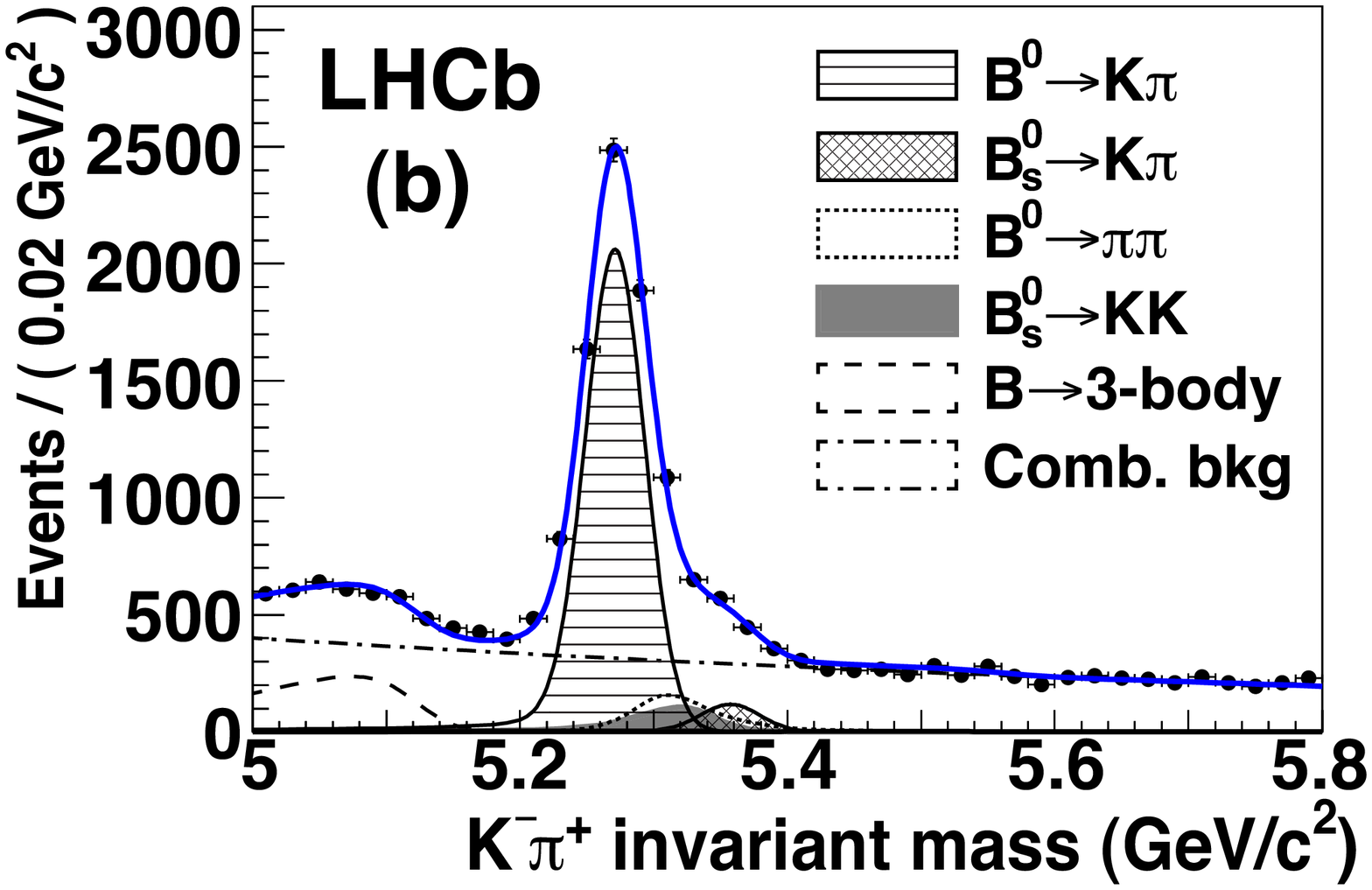}
\includegraphics[width=5.cm]{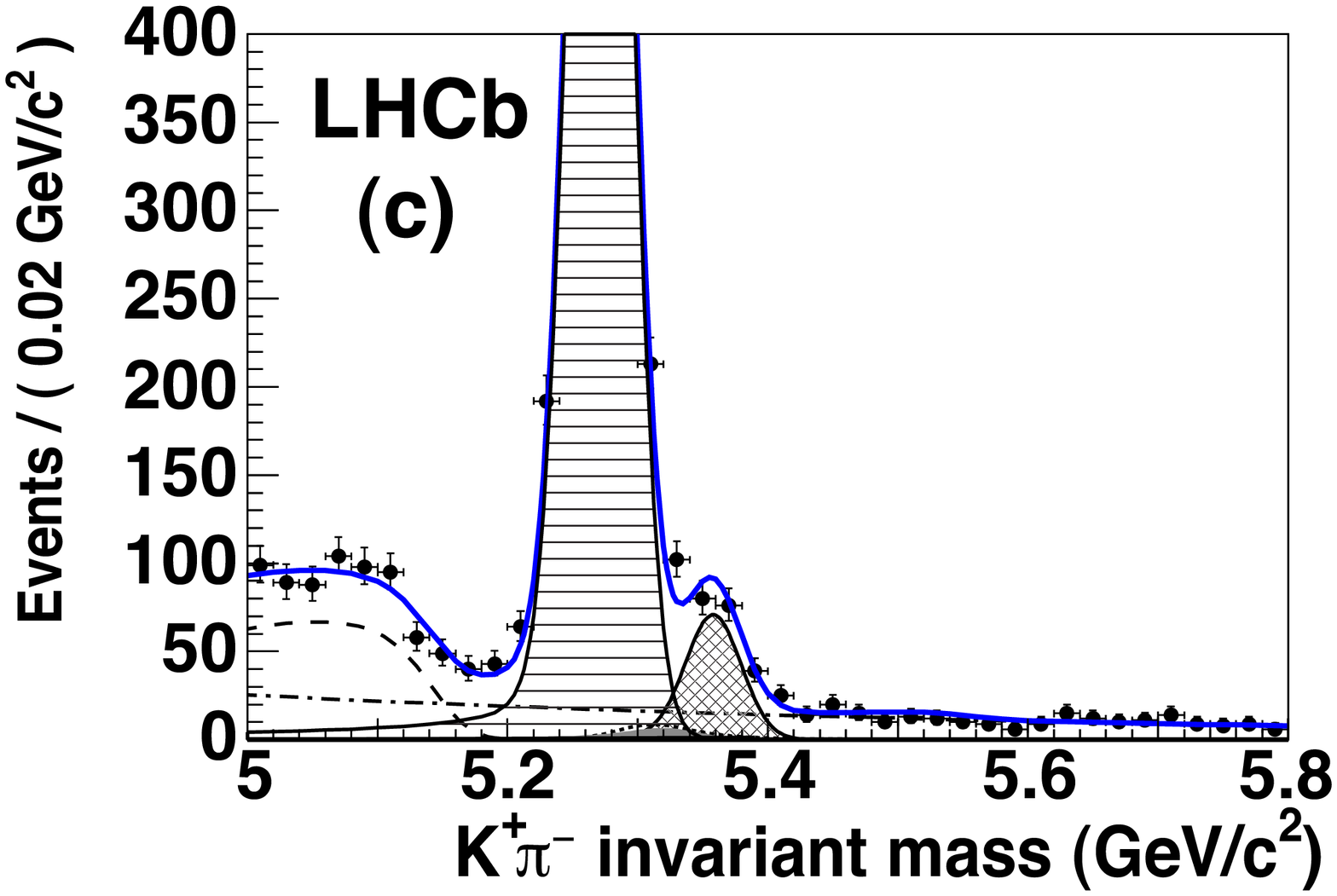}
\includegraphics[width=5.cm]{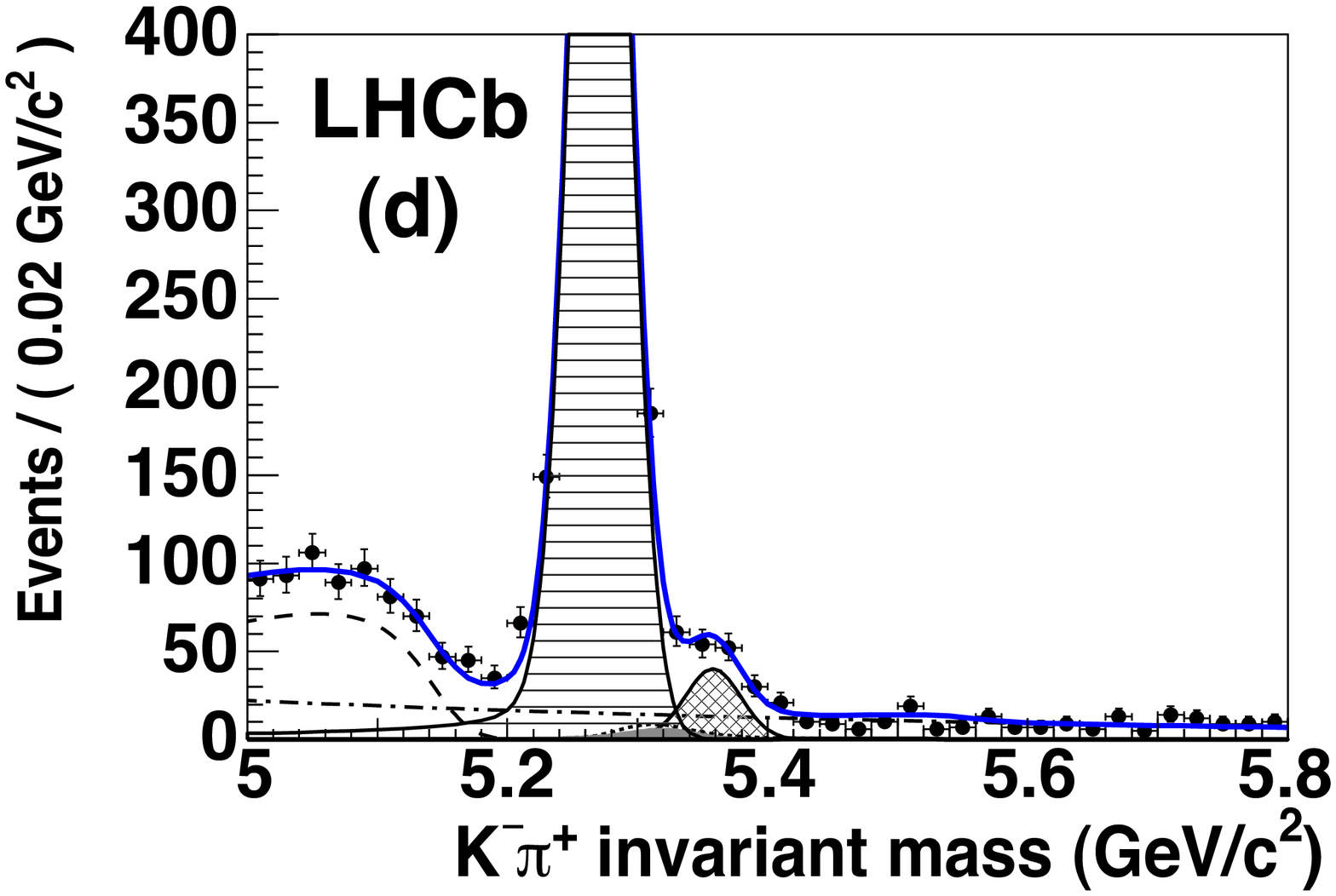}
\caption{\small Invariant $K\pi$ mass spectra obtained using the event selection adopted for best sensitivity 
to (a),(b)\acp(\Bz) and (c),(d)\acp(\Bsz). Plots (a) and (c) represent the \Kp\pim invariant mass whereas plots (b) and (d)
represent the \Km\pip invariant mass. The results of unbinned maximum likelihood fits are overlaid. The main components
contributing to the fit model are also shown.}
\label{fig:bhh}
\end{figure}

The probability for a \bquark quark to decay as \Bsz is  about 14 times smaller
than that to \Bz, and a stronger rejection of the combinatorial background was required.
The polarity of the LHCb magnetic field is reversed from time to time in order to partially cancel the effects of instrumental
charge asymmetries. Of the 0.35~fb$^{-1}$, about 43\% was acquired with one polarity and the rest with  opposite polarity.
The LHCb hadronic trigger is particularly efficient for the two-body modes. It requires one
very high transverse momentum good quality track in the event.
To distinguish the $\kaon \pi$ final
states we rely on the RICH particle identification system. The efficiencies and misidentification rates were carefully controlled
using large calibration data samples of $D^*\to D\pi\to(K\pi)_D \pi$ and   $\Lambda _b\to p\pi$ decays. The raw asymmetry is computed from maximum likelihood fits to the
$\kaon \pi$ invariant mass spectra shown in Figure~\ref{fig:bhh}. The raw asymmetry needs to be corrected for two residual effects.
The first is induced  by the detector acceptance, event reconstruction and the strong interaction of the final state particles with
the detector material. The second is associated with the production of the $B$ mesons and takes into account the
dilution factor due to mixing. The overall correction to the raw asymmetry is $\Delta\acp(\Bd)=-0.007\pm 0.007$ and
$\Delta\acp(\Bs)=0.010\pm 0.002$ , where the errors are statistical. The
final measured \CP asymmetries are
\begin{eqnarray}
\acp(\Bsz)=0.27\pm 0.08\stat \pm 0.02 \syst \, , \nonumber \\
\acp(\Bz)=-0.088\pm 0.011 \stat \pm 0.008 \syst \, , \nonumber \\
\end{eqnarray}
where the systematic uncertainties are dominated by the  correction effects and in the case of the  \Bsz there is also a leading
contribution from the combinatorial background description.

\section{Direct \CP violation in \kpipip and \kkkp decays}

With relatively large branching fractions of $\cal{O}$$(10^{-5})$ the  \kpipip and \kkkp  signals have comparable
data samples and the analysis differs only in the particle identification 
and in specific peaking background estimation.
The odd number of kaons in the final state and kinematics similarity enable a straight forward use of the large  \jpsik control
sample to account for residual
effects from the production mechanism, as well as the differences in the \Kp  and \Km interaction with the detector material.
The raw charge asymmetry, \acpraw, calculated from the positive and negative event yields and the corrected \acp is given by
$\acp(\Kpm h^+h^-) = \acpraw(\Kpm h^+h^-) - \acpraw(\jpsi \Kpm) + \acp(\jpsi \Kpm)$ where  $hh$ stands for $KK$ or $\pi\pi$ and the current
PDG value for $\acp(\jpsi \Kpm) $ is $(1\pm7)\times 10^{-3}$~\cite{Nakamura:2010zzi}. The subtraction was done separately in sub samples
 of the data divided by trigger selections, in order to cancel a possible trigger asymmetry for kaons. The total event
  yields are listed in 
Table~1 determined from unbinned extended
maximum likelihood fits to the $B^{\pm}$ candidate invariant
masses in the range $5079-5479\mevcc$.

\begin{table}[b]
\begin{center}
\begin{tabular}{l|ccccc}
\hline
mode          & \fskpipi         & \fskkk         & \fsjpsik & \fspipipi         & \fskkpi\\  \hline
$B^{-}$    & $18\,168\pm170$  & $10\,289\pm110$  & $30\,140\pm179$ & $\valbmpipipi\pm\errbmpipipi$  & $\,\;\valbmkkpi\pm\errbmkkpi$  \\
$B^{+}$    & $17\,540\pm169$  & $11\,606\pm117$  & $30\,984\pm182$ & $\valbppipipi\pm\errbppipipi$   & $\,\;\valbpkkpi\pm\errbpkkpi$ \\
\hline
\end{tabular}
\caption{Event yields for $B^{-}$ and $B^+$ samples and for the whole data set. }
\end{center}
\label{fitRes}
\end{table}

The final results for the inclusive charge asymmetries of \kpipi and \kkk are:
\begin{eqnarray}
 \acp(\kpipi) =   +0.034 \pm 0.009\stat \pm 0.004\syst \pm 0.007(J/\psi \Kpm) \, , \nonumber \\
 \acp(\kkk)   =  -0.046 \pm 0.009\stat \pm 0.005\syst \pm 0.007(J/\psi \Kpm) \, , \nonumber
\end{eqnarray}
where the first uncertainty is statistical, the second is systematic, and the third is the uncertainty of the \acp of \jpsik from the PDG~\cite{Nakamura:2010zzi}.
The main contributions to the systematic uncertainty are attributed to the use of the control channel to account for residual asymmetries to
 signal model and to the
acceptance in the Dalitz plot.

\begin{figure}[tb]
\centering
\includegraphics[width=5.cm]{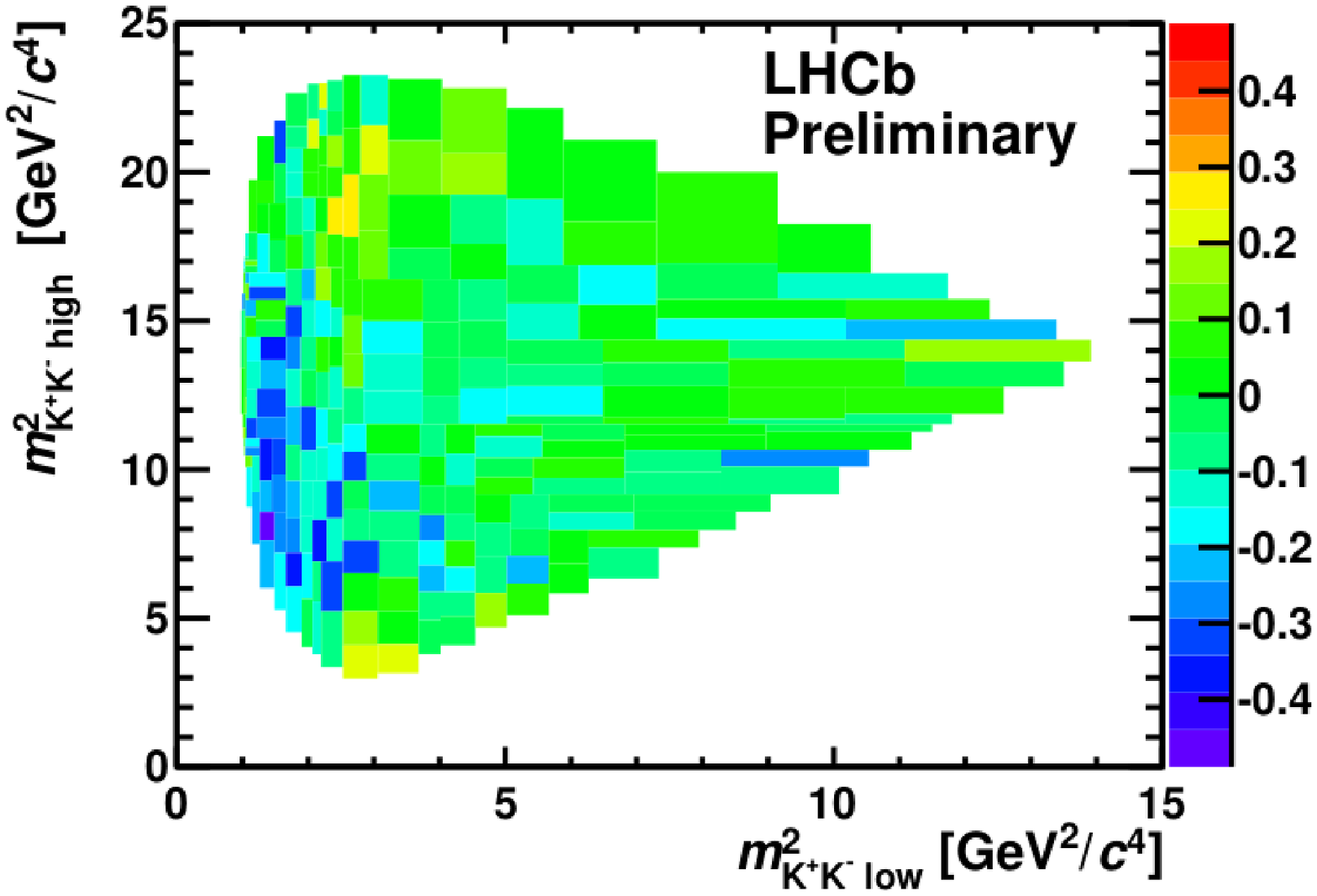}
\includegraphics[width=5.cm]{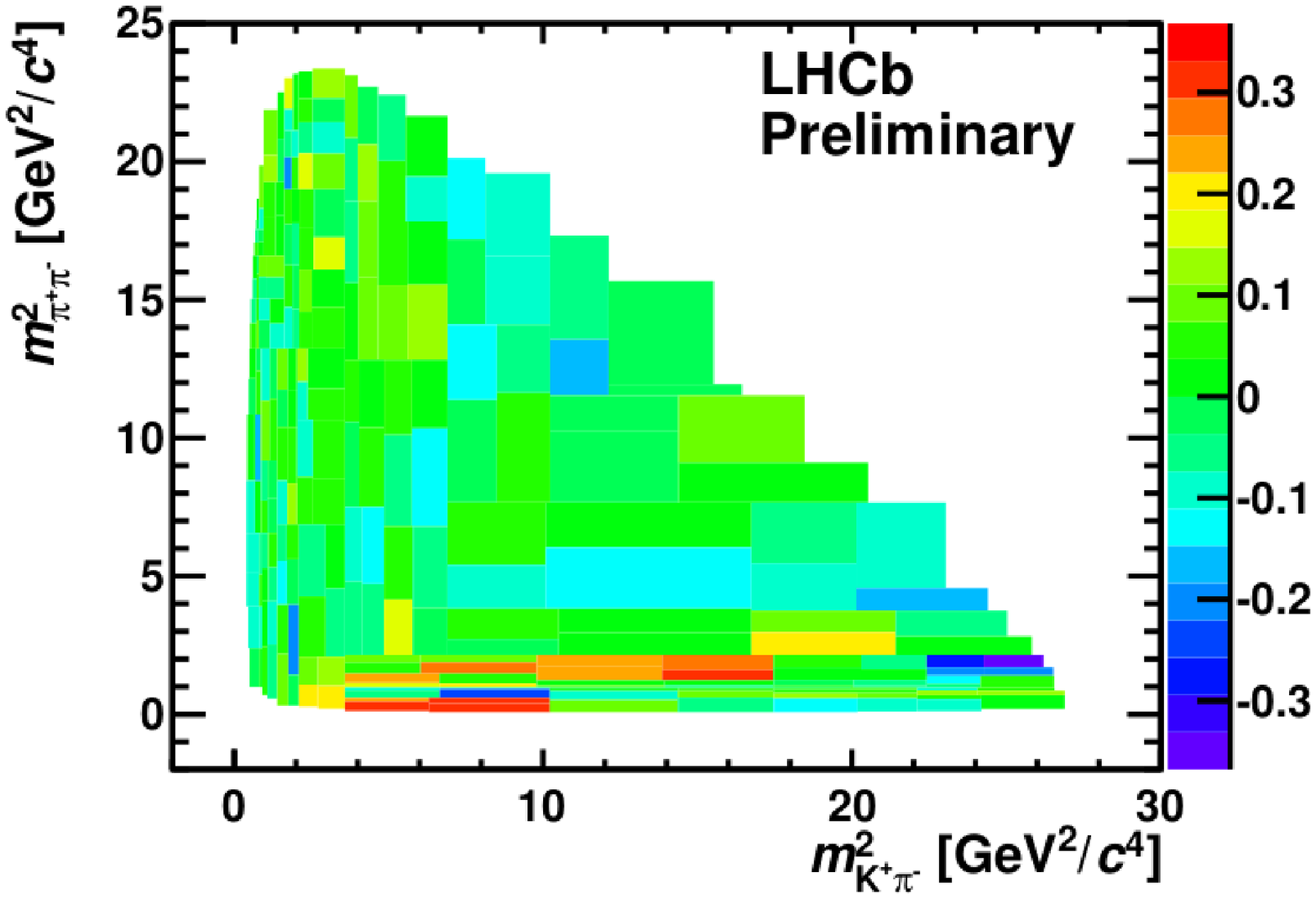}
\includegraphics[width=5.cm]{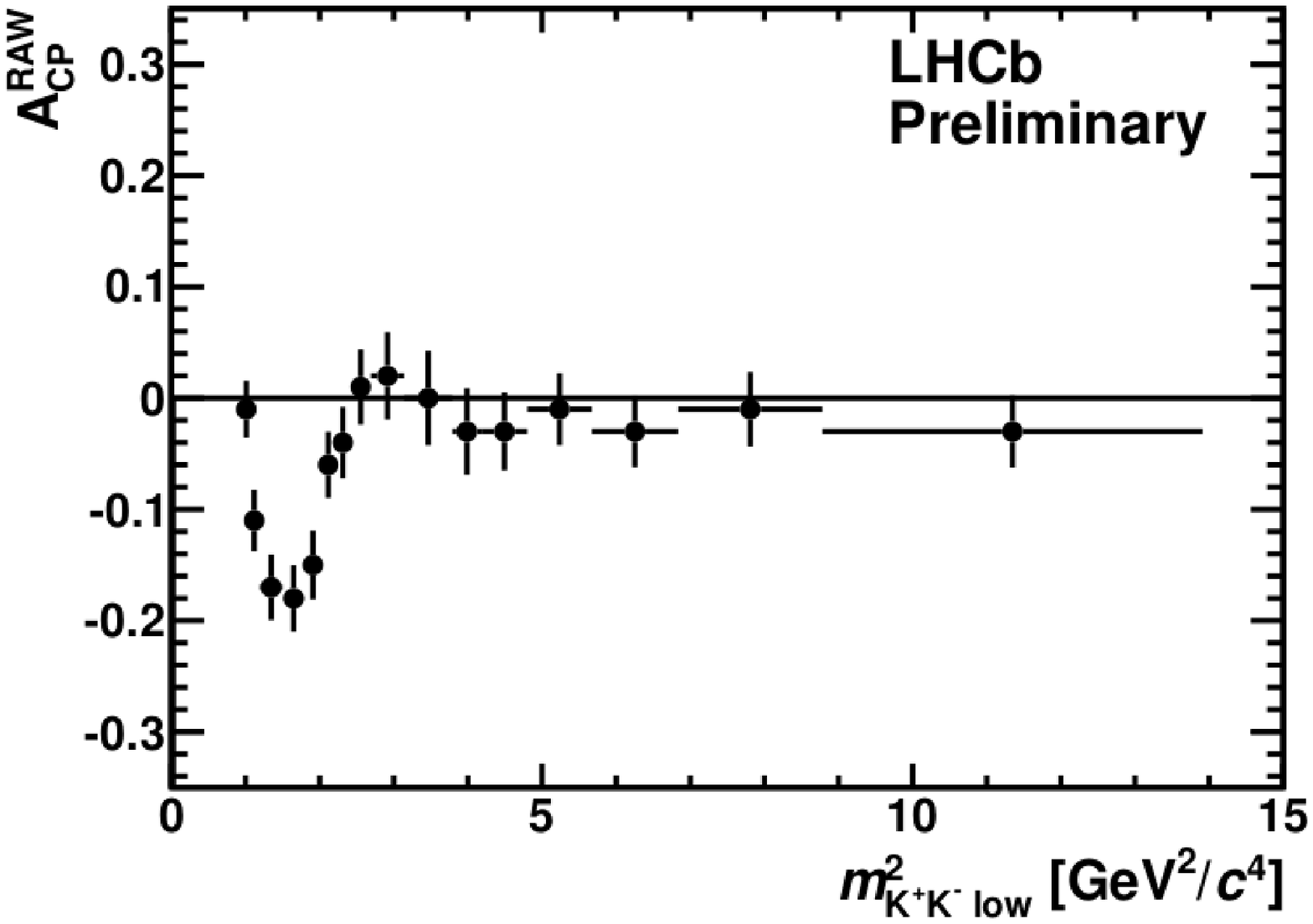}
\includegraphics[width=5.cm]{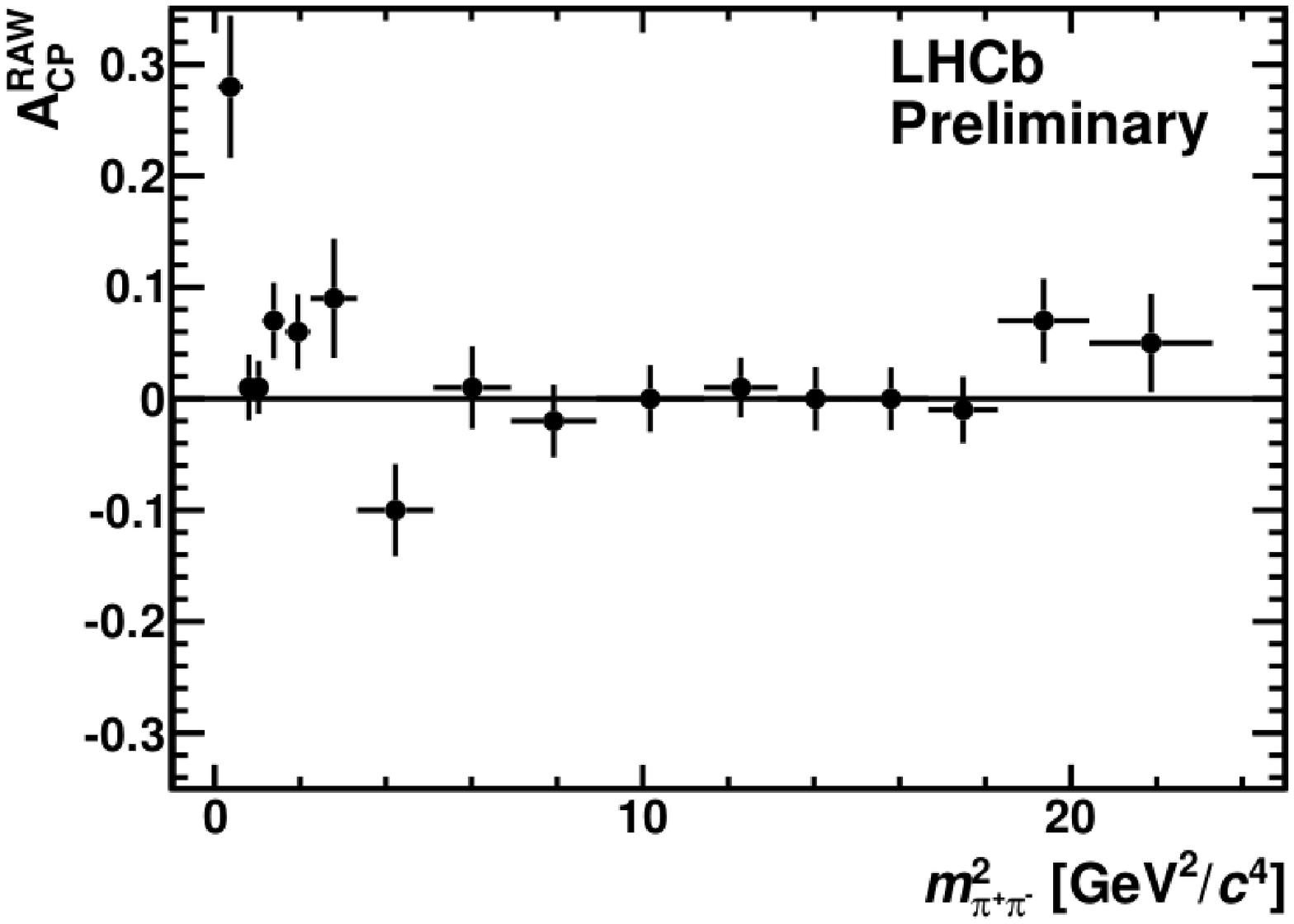}
\caption{\small \acpn in Dalitz plot bins with equal event populations for \kkk (top left) and \kpipi (top right). 
The asymmetry \acpn is calculated from the bin contents including backgrounds, and is not corrected for 
production or detection asymmetries. \acpraw  for $m^2_{\Kp \Km\,{\rm low}}$ (bottom left) and $m^2_{\pip \pim}$ (bottom right) projections.}
\label{fig:acpdalitzkpipikkk}
\end{figure}

The two relevant variables used to represent the Dalitz plot are the two-body invariant masses ($m^2_{\pip \pim}$, \mmkpi)
in the case of \kpipi decay and ($m^2_{\Kp \Km\,{\rm low}}$, $m^2_{\Kp \Km\,{\rm high}}$)
in the case of \kkk decay, where the symmetrical phase space is folded.
To visualise the asymmetries in phase space, the Dalitz plots of candidates with three-body invariant masses within
$\pm 40\mevcc$ of the peak were divided into bins with equal population 
using an adaptive binning algorithm.
In each bin, the variable $\acpn\equiv \frac{N^- - N^+}{N^- + N^+}$  was computed from the number of \Bpm event candidates, $N^{\pm}$,
which includes signal and background events in that bin and is not corrected for efficiency.
The resulting \acpn distributions over the Dalitz plots are shown in the upper plots of Figure~\ref{fig:acpdalitzkpipikkk}.
Large asymmetries are clearly visible in  the low \mmkklow and low \mmpipi regions respectively for  \kkk and \kpipi.

To analyse the \CP asymmetries over the phase space without background influence, the raw asymmetries were plotted in
bins of the two-body invariant mass projections, shown in the bottom plots of Figure~\ref{fig:acpdalitzkpipikkk}. We observe large opposite sign
asymmetries in the low \mmpipi and \mmkklow invariant masses not clearly associated to a resonant state. 

\section{Direct \CP violation in \kkpip and \pipipip decays}
The samples of \kkpip and \pipipip are of comparable sizes, much smaller than the \kpipip and \kkkp ones and with much higher background levels.
We measure the raw asymmetry defined in terms of the \Bp and \Bm event yields and corrected by the acceptance in the Dalitz plot as
$\acprawacc = \frac{(N_{\Bm}/R) - N_{\Bp}}{(N_{\Bm}/R) + N_{\Bp}}$, where $R$ is the ratio between the \Bp and \Bm data-weighted
average efficiency in the Dalitz plot, found to be $R_{\pi\pi\pi} = 1.023\pm 0.029$ and $R_{KK\pi} = 0.991\pm 0.027$. The total signal yields 
 are listed in Table~1.

Here the net strangeness in the  final state is zero  and the residual detection asymmetry refers to a pion and is
measured by LHCb to be  $\adetpi=(0.00\pm 0.25)\%$\cite{Aaij:2012cy} using large  $D^{*\pm}$  calibration samples. The residual production asymmetry is  estimated from the
\jpsik control channel, eliminating
the factor relative to the detection associated to bachelor kaon  measured from calibration samples, to be $\adetk=-0.010\pm0.002$\cite{Aaij:2012qe}.   

In order to cancel out a possible trigger asymmetry for hadrons, the final asymmetries are calculated from data  
divided into two subsamples depending on the trigger selection.
The inclusive \CP asymmetries are the weighted averages of the trigger subsamples and are measured to be
\begin{eqnarray}
\acp(\pipipi)= +\valacppipipiAcc \pm \errstatacppipipiAcc \stat \pm  \errsystacppipipiAcc \syst \pm \erracpjpsikpdg(\jpsi\Kpm) \, , \nonumber \\
\acp(\kkpi)= \valacpkkpiAcc \pm \errstatacpkkpiAcc \stat \pm \errsystacpkkpiAcc \syst  \pm \erracpjpsikpdg(\jpsi\Kpm) \, , \nonumber
\end{eqnarray}
where the first uncertainty is statistical, the second is the experimental systematic uncertainty, and the third is the
uncertainty of the \acp value of \jpsik from the PDG~\cite{Nakamura:2010zzi}.

\begin{figure}[htb]
\begin{center}
\includegraphics[width=4.5cm]{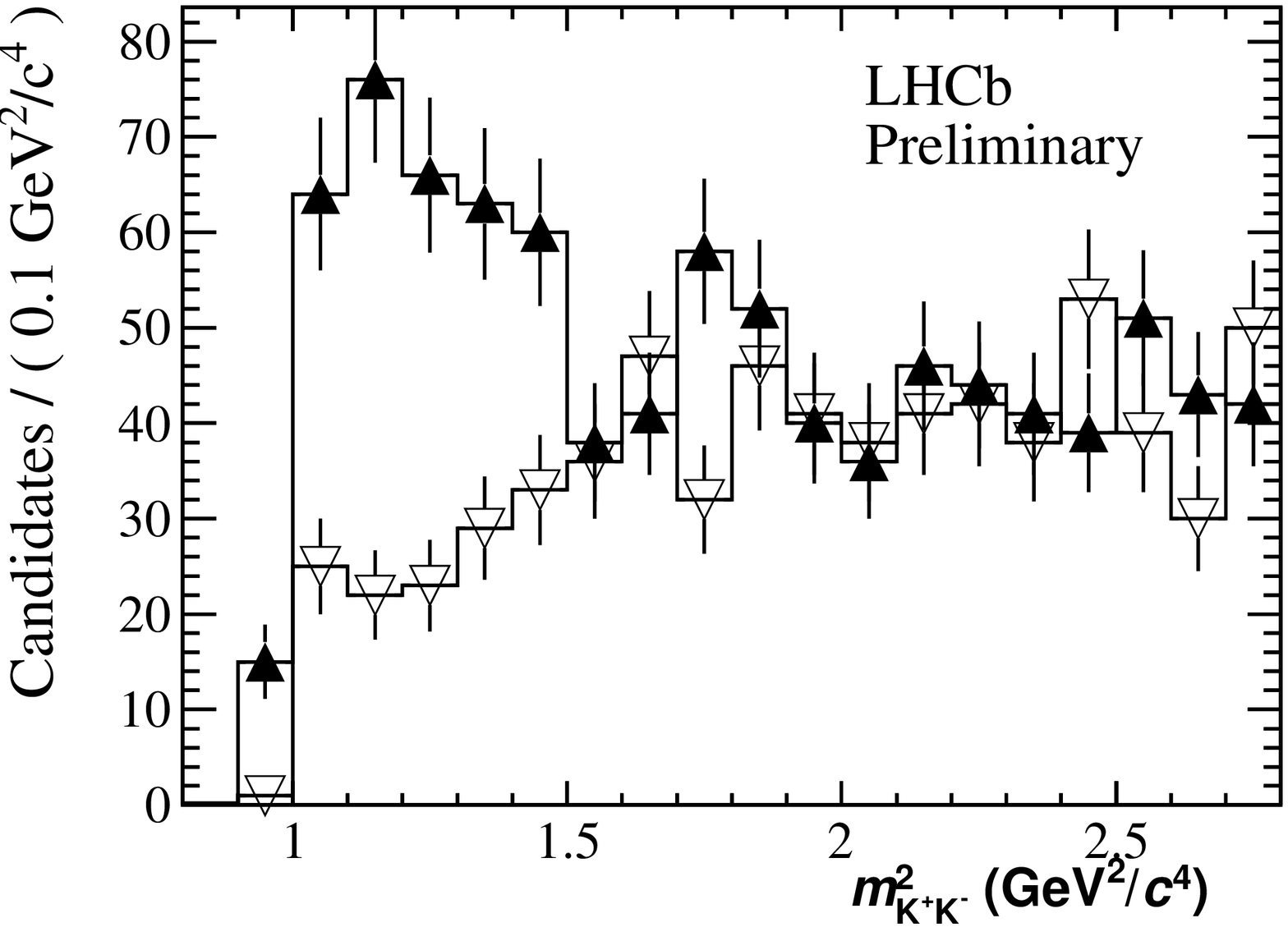}
\includegraphics[width=4.5cm]{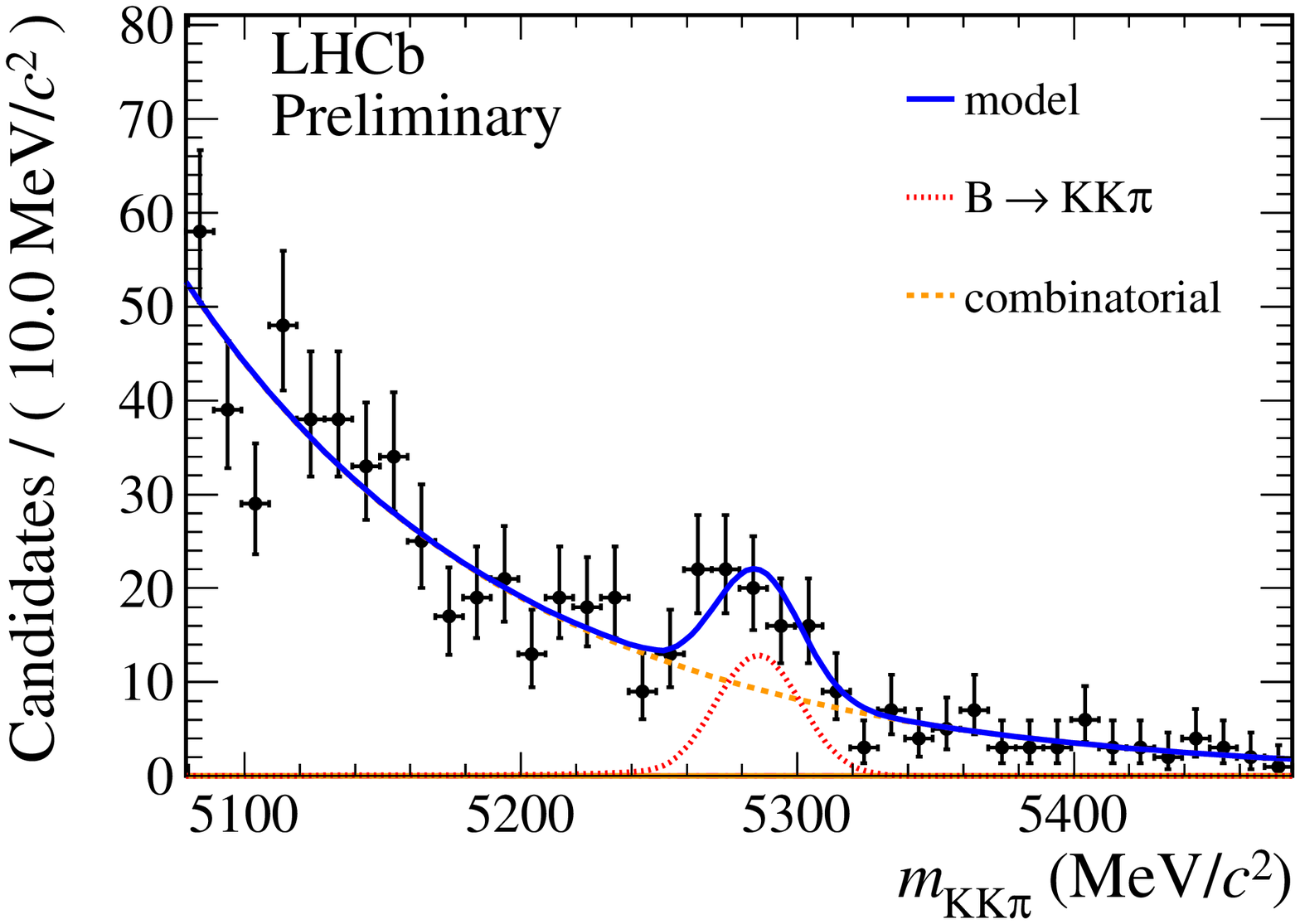}
\includegraphics[width=4.5cm]{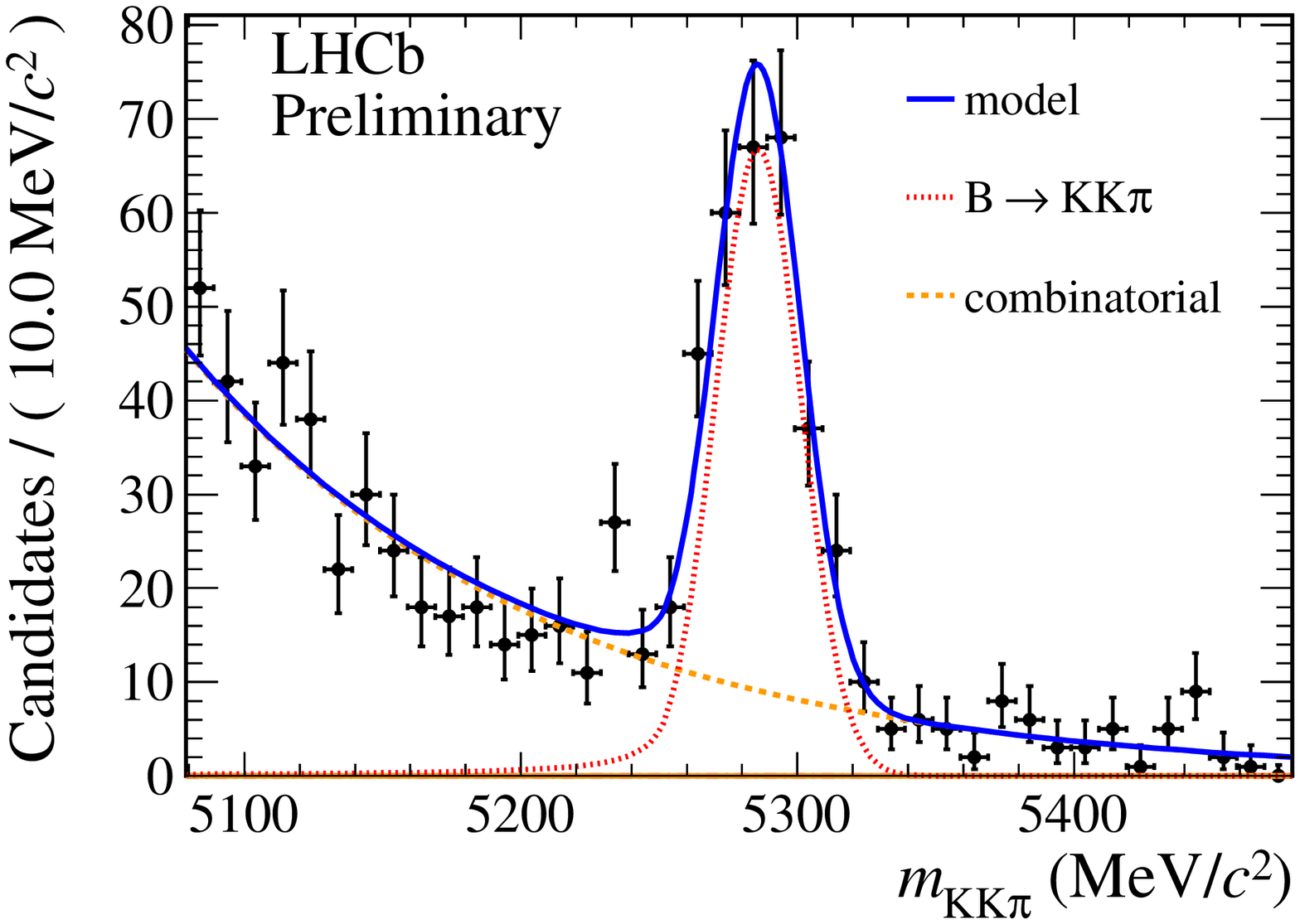}
\caption{\kkpi event yields (signal and background) as a function of \mmkk (left), where the empty triangles represent \Bm and
the filled triangles represent \Bp; fits of \Bm (center) and \Bp (right) for candidates with  $\mmkk<1.5\gevgevcccc$. }
\label{fig:StructKKPlot_KKPi}
\end{center}
\end{figure}

\begin{figure}[htb]
\begin{center}
\includegraphics[width=4.5cm]{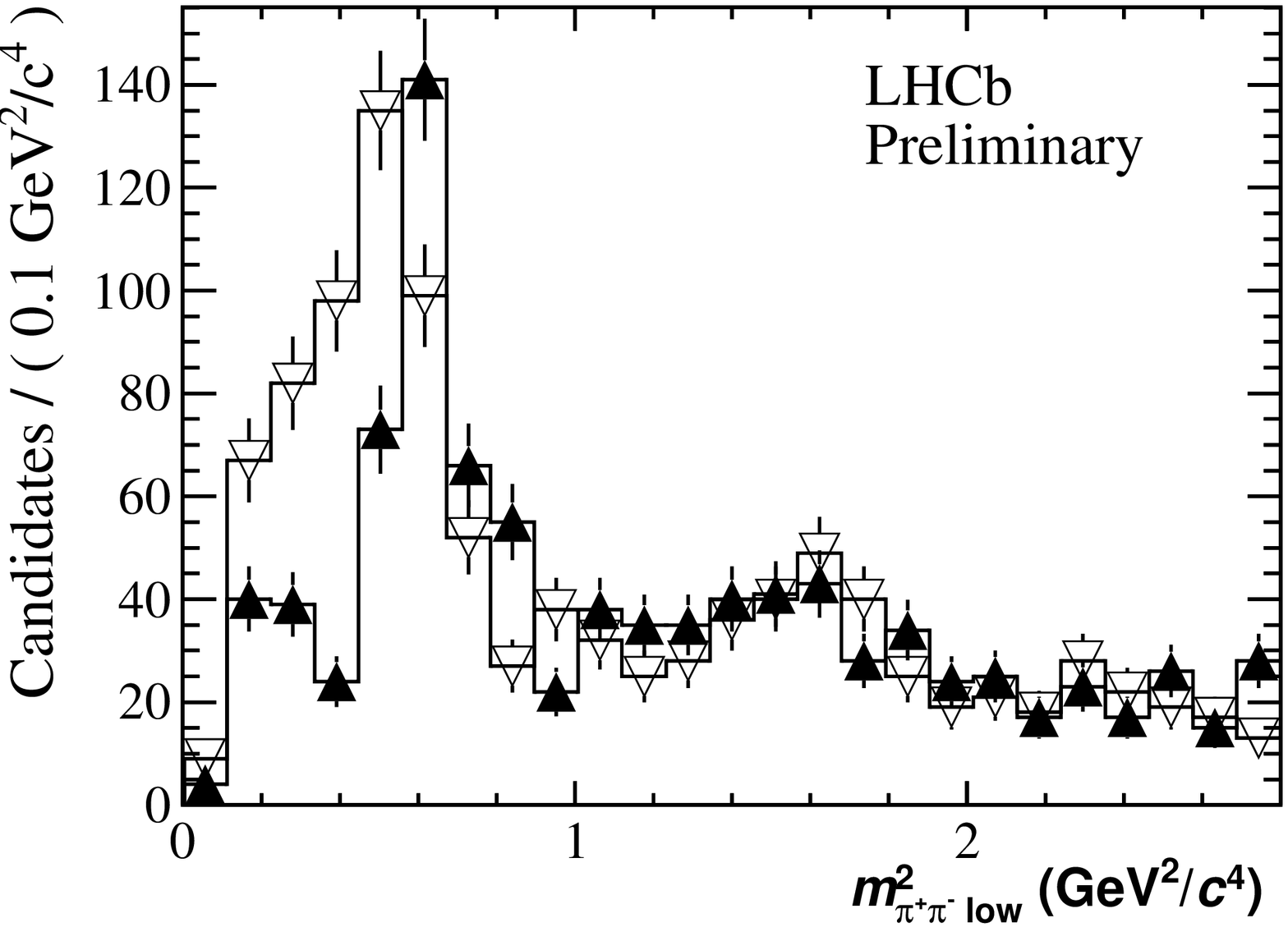}
\includegraphics[width=4.5cm]{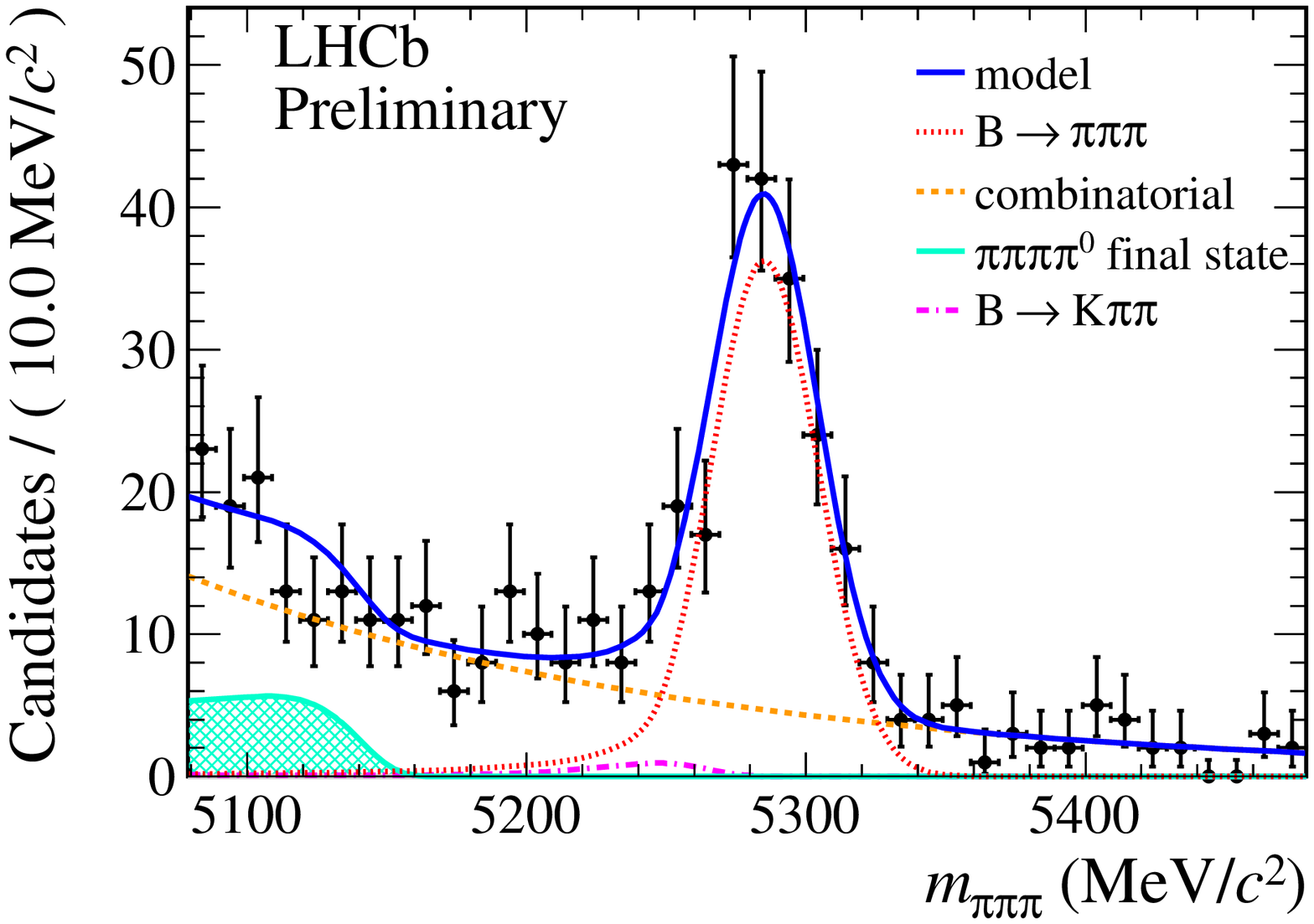}
\includegraphics[width=4.5cm]{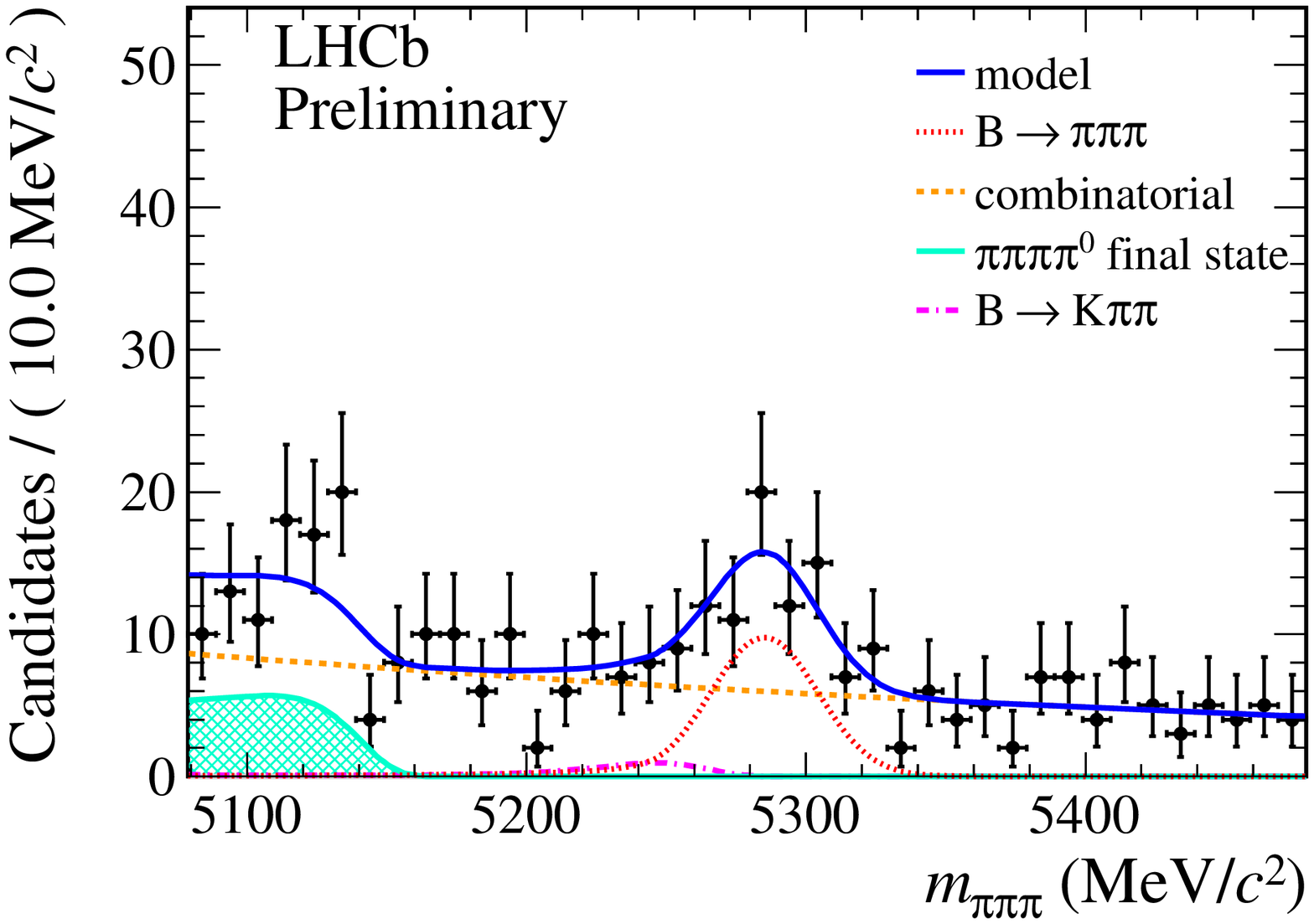}
\caption{\pipipi event yields (signal and background) as a function of \mmpipilow for $\mmpipihi>15\gevgevcccc$ (left), where the
empty triangles represent \Bm and the filled triangles represent \Bp; and fits of \Bm (center) and \Bp (right) for candidates with $\mmpipilow<0.4\gevgevcccc$ and $\mmpipihi>15\gevgevcccc$. 
}
\label{fig:StructPlot_PiPiPi_massFit}
\end{center}
\end{figure}

Investigating the distribution of asymmetries in the Dalitz plots we observe large negative contributions at low \mmkk invariant mass as well
as large positive asymmetries at low \mmpipi, particularly in the high  \mmpipihi region, as can be seen in Figures~\ref{fig:StructKKPlot_KKPi}
and \ref{fig:StructPlot_PiPiPi_massFit}. To quantify this effect we compute the local \CP asymmetry in
the phase space regions  in the same way as the inclusive one, correcting
for the acceptance, production asymmetry and detection asymmetry.
The  \CP asymmetry of \pipipi in the  region $\mmpipilow < 0.4\gevgevcccc$ and $\mmpipihi > 15\gevgevcccc$ is
determined to be
\begin{eqnarray}
\acp(\pipipi\;{\rm region})= +\valacppipipiregAcc \pm \erracppipipiregAcc \stat \pm  \errsystacppipipiregAcc \syst \pm
\erracpjpsikpdg(\jpsi\Kpm) \, . \nonumber
\end{eqnarray}
The \CP asymmetry of \kkpi in the  region $\mmkk< 1.5\gevgevcccc$ is measured to be
\begin{eqnarray}
\acp(\kkpi \; {\rm region})= \valacpkkpiregAcc \pm \erracpkkpiregAcc \stat \pm \errsystacpkkpiregAcc \syst  \pm \erracpjpsikpdg(\jpsi\Kpm) \, . \nonumber
\end{eqnarray}
 
\section{Conclusion}

We have observed a rich pattern of \CP violation in the charmless decays of the \B mesons.
Contrary to  naive expectations, we observe large \CP violation in the \kkpip, \pipipip. Intriguingly large
opposite sign \acp is observed in the low $\Kp\Km$ and $\pip\pim$   invariant mass regions of the three-body decays. 
The effect  seems not to be associated to any resonat state. LHCb  will triple its sample with the data collected in 2012.

\def\Discussion{
\setlength{\parskip}{0.3cm}\setlength{\parindent}{0.0cm}
     \bigskip\bigskip      {\Large {\bf Discussion}} \bigskip}
\def\speaker#1{{\bf #1:}\ }
\def\endDiscussion{}
\end{document}